\documentclass[12pt]{iopart}

\newcommand{\beq}{\begin{equation}}
\newcommand{\eeq}{\end{equation}}

\newcommand{\eqref}[1]{(\ref{#1})}

\begin{document}

\title{A local induced action for the noncritical string}

\author{Willem Westra$^1$ and Stefan Zohren$^2$} 

\address{ $\!^1$ Division of Mathematics, University of Iceland, Reykjavik, Iceland } 
\address{$\!^2$ Rudolf Peierls Centre for Theoretical Physics, 1 Keble Road, Oxford OX1 3NP, UK} 
\ead{\mailto{wwestra@raunvis.hi.is}, \mailto{zohren@physics.ox.ac.uk}}

\begin{abstract}
We present an alternative to Polyakov's induced action for the noncritical string. Our Yang-Mills like action is both local and invariant under coordinate transformations.  It defines a teleparallel theory of gravity with interesting links to Ho\v{r}ava-Lifshitz and Einstein-aether theories. 
It is of Liouville type in the conformal gauge while, remarkably, in  the proper-time gauge it gives an effective action familiar from the Causal Dynamical Triangulation (CDT) approach to 2d quantum gravity. In the latter gauge the effective action is especially interesting since its quantization is known to reduce to a quantum mechanical model. 
\end{abstract}

\maketitle

\section{The local effective action}
As in the original computation \cite{Polyakov:1981rd} due to Polyakov\footnote{For a textbook treatment see \cite{Hatfield:1992rz}.}  we formally define the quantum theory of noncritical strings by a path integral over metrics modulo diffeomorphisms and a set of scalar fields which represent the target space coordinates, 
\beq \label{formalPI}
Z = \int \mathcal{D} g_{\alpha\beta} \mathcal{D} X^I e^{-S_{p}(g_{\alpha\beta}, X^I)}, 
\eeq
where $S_{p}$ is the Polyakov surface action with an additional bare cosmological term $\mu_b$ on the world sheet\footnote{We consider a fixed topology of the worldsheet and therefore neglect the Einstein-Hilbert term of the action.},
\beq
S_{p} (g_{\alpha\beta}, X^I) =  \frac{1}{2}\int d^2x \sqrt{g}\left( g^{\alpha \beta} \partial_\alpha X^I \partial_\beta X_I+\mu_b \right).
\eeq
After performing the path integral over the quantum fluctuations of the scalar fields one obtains the standard formula for the induced action,
\beq
S_{ind} = \frac{d}{2} \left[ \log\mbox{det}\Delta\right] =  \frac{d}{2} \left[ \mbox{Tr}(\log\Delta)\right],
\eeq
which we determine\footnote{We gloss over issues with zero modes and divergencies which can be absorbed into Newton's constant to keep the main point of our discussion as clear as possible.}, as usual, by deriving a differential equation for it, 
\beq \label{varlap}
\delta S_{ind} = -\frac{d}{2} \int_\varepsilon^\infty ds
 \int d^2 z \sqrt{g}\  \delta \Delta\langle z| e^{-s \Delta} |z\rangle.
\eeq
At this point we differ slightly from the standard presentation, see for instance \cite{Alvarez:1982zi}. Instead of considering a Weyl variation of the metric we consider a variation of the square root of the determinant of the metric. The Laplacian can be written as follows,
\beq
\Delta = -\frac{1}{\sqrt{g}}\partial_\alpha(\tilde{g}^{\alpha\beta} \partial_\beta),
\eeq
where we have introduced the decomposition $g_{\alpha\beta} = \sqrt{g}\frac{g_{\alpha\beta} }{\sqrt{g}}= \sqrt{g}\tilde{g}_{\alpha\beta}$. The object $\tilde{g}_{\alpha\beta}$ is a metric density of weight $-1$ with unit determinant. It transforms under coordinate transformations as follows,
\beq
\tilde{g}_{\alpha\beta}=
 \left|\frac{\partial x'^\gamma}{\partial x^\delta}\right|^{-1}\frac{\partial x^{\alpha'}}{\partial x^{\alpha}} \frac{\partial x^{\beta'}}{\partial x^{\beta}}\tilde{g}_{\alpha' \beta'}.
\eeq
This transformation law is better known in its infinitesimal form where it defines the ``conformal Killing operator'' $\delta_{v}\tilde{g}_{\alpha \beta} =\frac{1}{\sqrt{g}} (Pv)_{\alpha\beta}$,
\beq \label{pop}
\frac{1}{\sqrt{g}} (Pv)_{\alpha\beta}
=\frac{1}{\sqrt{g}}\left( \nabla_\alpha v_\beta + \nabla_\beta v_\alpha-  g_{\alpha\beta} \nabla^\gamma v_\gamma\right).
\eeq
The determinant of this operator is related to the ghost sector of the gravitational part of the path integral, in this paper however we restrict our attention to the matter sector.
We choose arbitrary fluctuations which only affect the determinant of the metric and therefore preserve the metric density  $\tilde{g}_{\alpha\beta}$. The variations in $\tilde{g}_{\alpha\beta}$ are not necessary to obtain the induced action, the conformal variations and covariance arguments turn out to be sufficient.  As in the case of Weyl fluctuations, the variation of the Laplacian is then  proportional to itself,
\beq \label{flucdelta}
\delta\Delta = (\delta\sqrt{g})\frac{\partial\Delta }{ \partial \sqrt{g}} = -\frac{\delta\sqrt{g}}{\sqrt{g}}\Delta.
\eeq
Using \eqref{flucdelta}, the definition for the Green's function and its differential equation,
\beq
G(z,z';s) = \langle z| e^{-s \Delta} |z'\rangle,
\eeq
\beq
(\partial_s + \Delta)G =0,
\eeq
and its expansion for infinitesimal $s$, see \cite{Alvarez:1982zi}, and inserting in \eqref{varlap} one obtains
\beq \label{varS}
\delta S_{ind} = -\frac{d}{2} \lim_{\varepsilon \rightarrow 0}\left[
 \int d^2 z \delta \sqrt{g} \  \left(\frac{1}{4\pi \varepsilon} + \frac{1}{24 \pi} R\right) \right].
\eeq
Contact with Liouville theory is made by identifying the logarithm of the square root of the metric determinant with the Liouville field,
\beq \label{phidef}
\phi(x) = \log \sqrt{g}.
\eeq
From this identification it should be clear that in our conventions \emph{the Liouville field is not a scalar field} but transforms additively under general coordinate transformations with the corresponding Jacobian determinant,
\beq \label{jacob}
\phi(x'(x)) = \phi(x) - \log \left|\frac{\partial x'^{\alpha}}{\partial x^\beta}\right|,
\eeq
which implies that its exponential is a scalar density of weight one.
Using \eqref{phidef} we write the variational equation for the induced action \eqref{varS} as a differential equation in terms of the Liouville field,
\begin{equation} \label{dliouville}
\delta S_{m}\! \!=\!- \frac{d}{48 \pi} \!
\! \int \! d^2 z\delta \phi\! \left(\!\tilde{R}\!-\partial_\alpha (\tilde{g}^{\alpha \beta} \partial_\beta \phi )\!+\!\mu e^{\phi}
\right),
\end{equation}
where $\tilde{R}$ is the scalar curvature of $\tilde{g}_{\alpha\beta}$ and $\mu=6/\varepsilon$. Integrating the above with respect to $\phi$ we obtain an action which is given by,
\beq \label{stotdef}
S_{m} = -\frac{d}{48\pi} \left(S_L +  \int d^2 z  f(\tilde{g}_{\alpha\beta})\right),
\eeq
where $S_L$ is a Liouville type action, 
\begin{equation}  \label{sliou}
S_L=
 \int d^2 z  \ \Big(\frac{1}{2}\tilde{g}^{\alpha \beta} \partial_\alpha \phi \partial_\beta \phi + \phi \tilde{R} +\mu e^\phi \Big).
\end{equation}
Note that the second term in \eqref{stotdef} can be seen as a consequence of the fact that one only considers variations with respect to $\sqrt{g}$ in \eqref{varS}.
We emphasize that the effective action \eqref{stotdef}, \eqref{sliou} is derived without fixing a gauge.
However, neither the Liouville field nor $\tilde{R}$ are scalars, and $\tilde{g}_{\alpha\beta}$ is not a tensor therefore \emph{the Liouville action \eqref{sliou} is not invariant under coordinate transformations}. We thus need a specific term $f(\tilde{g}_{\alpha\beta})$ in \eqref{stotdef} to obtain a reparametrization invariant effective action. 

So far the results are relatively standard but from this point onward we differ in a pivotal way from the conventional procedure to obtain a covariant induced action. We determine the action by making the following local and reparametrization invariant ansatz,
\beq \label{ansatz} 
S_{m} =  -\frac{d}{48\pi} 
\int d^2z \left(\frac{1}{2} 
\tilde{g}^{\alpha \beta}  \tilde{D}_\alpha \phi \tilde{D}_\beta \phi
+\mu e^\phi
\right), 
\eeq
where a covariant derivative is introduced with a ``conformal connection''\footnote{Note that our connection is introduced to compensate for conformal coordinate transformations not Weyl transformations. For a formally similar gauging of Weyl transformations see \cite{O'Raifeartaigh:1996hf}, which is based on Weyl's original ideas, see also \cite{Jackiw:2005su}.}  which compensates the additive behavior of $\partial_\alpha \phi$ under coordinate transformations \eqref{jacob},
\beq
 \tilde{D}_\alpha \phi = \partial_\alpha \phi - \tilde{\omega}_\alpha.
\eeq
By partial integration we obtain
\begin{eqnarray}
&S_{m} &= -\frac{d}{48\pi} \int d^2z \Big( \frac{1}{2}\tilde{g}^{\alpha \beta} \partial_\alpha \phi \partial_\beta \phi + \label{sexpl} \\ 
&&
+\phi\partial_\alpha( \tilde{g}^{\alpha \beta} \tilde{\omega}_\beta) 
+\frac{1}{2}\tilde{g}^{\alpha \beta} \tilde{\omega}_\alpha \tilde{\omega}_\beta+ \mu e^\phi 
\Big)\nonumber.
\end{eqnarray}
If we compare to \eqref{stotdef} and \eqref{sliou} one obtains the following two equations for the conformal connection $\tilde{\omega}_\alpha$ and the function $f(\tilde{g}_{\alpha\beta})$,
\begin{eqnarray}
\partial_\alpha (\tilde{g}^{\alpha \beta} \tilde{\omega}_\beta) &= \tilde{R} \label{confcondef} ,\\
 f(\tilde{g}_{\alpha\beta})&= \tilde{g}^{\alpha \beta} \tilde{\omega}_\alpha \tilde{\omega}_\beta. 
\end{eqnarray}
From \eqref{confcondef} one can see that $\tilde{\omega}_\alpha$ does not depend on the Liouville field $\phi$.
The conformal connection can be obtained by the observation that, see appendix, the Einstein-Hilbert Lagrangian in terms of vielbein variables is a total derivative,
\begin{eqnarray}
\sqrt{g}R &=\tilde{R} - \partial_\alpha (\tilde{g}^{\alpha \beta} \partial_\beta \phi)\label{rtilde},\\
&= \partial_\alpha \left(2\tilde{\varepsilon}^{\alpha \beta} \omega_\beta \right) \label{totderiv},
\end{eqnarray}
where the epsilon tensor $\varepsilon_{\alpha\beta}$ and the epsilon symbol $\tilde{\varepsilon}_{\alpha\beta}$ are related via $\varepsilon_{\alpha\beta}=\sqrt{g}\tilde{\varepsilon}_{\alpha\beta}$. Combining \eqref{confcondef}, \eqref{rtilde} and \eqref{totderiv}  gives
\beq
\tilde{\omega}_\alpha 
= 2\varepsilon_{\alpha}^{~ \beta} \omega_\beta + \partial_\alpha\phi,
\eeq
hence the conformal covariant derivative of the Liouville field is simply
\beq
 \tilde{D}_\alpha \phi = -2\varepsilon_{\alpha}^{~ \beta} \omega_\beta.
\eeq
Inserting this into \eqref{ansatz} gives,
\beq
S_{m} =  -\frac{d}{48\pi} 
\int d^2z \sqrt{g}
\left(
2g^{\alpha\beta}\varepsilon_{\alpha}^{~ \gamma} \varepsilon_{\beta}^{~ \delta} \omega_\gamma  \omega_\delta
+\mu\right).
\eeq
In terms of the fundamental fields, the vielbeins, this yields a remarkably simple form for the matter induced action which is local and manifestly invariant under coordinate transformations, 
\beq \label{resdef}
S_m = -\frac{d}{48\pi} S_F,
\eeq
where $S_F$ is the Yang-Mills like action,
\beq \label{result}
S_{F} =   
\int d^2z \sqrt{g}
\left(
F^a_{\alpha\beta} F^{\alpha\beta}_{a}+\mu
\right),
\eeq
and
\beq
F^a_{\alpha\beta} F^{\alpha\beta}_{a} = \delta_{ab}g^{\alpha\beta}g^{\gamma\delta}\partial_{[\alpha}e^a_{\gamma]}\partial_{[\beta}e^b_{\delta]}.
\eeq
As a check one can indeed see that the trace of the energy momentum tensor of \eqref{resdef} gives,
\beq
T^\beta_{~\beta}= -\frac{\pi}{e}e^a_\beta\frac{\delta S_m}{\delta e^a_{\beta}}= \frac{d}{12}\left(R + \frac{1}{2}\mu\right) =0.
\eeq
This is the expected result that our action represents an integrated form of the trace anomaly.
Note that the theory defined  by \eqref{resdef} should not be regarded as a gauge theory for local Lorentz transformations since the action breaks this symmetry. The action is however invariant under transformations of the following form
\beq
e^a_\alpha \rightarrow e^a_\alpha + \partial_\alpha \lambda^a
\eeq
which are sometimes interpreted as local translations. In general, theories that preserve this symmetry but violate local Lorentz invariance are called teleparallel theories for gravity \cite{Arcos:2005ec}. In this context the vielbeins can be thought of as gauge fields for the group of local translations and $F^a_{\alpha\beta}$ are the corresponding translational field strengths. General relativity is a special example of a teleparallel theory which is invariant both under local translations and local Lorentz transformations.  Our effective action however \eqref{result},  is an example of a teleparallel theory which
%
excites also the local rotational degrees of freedom of the gravitational field. These degrees of freedom are precisely the contributions which are contained in the vielbeins but not in the metric.
   
The action \eqref{result} should be contrasted with the non-local induced action introduced by Polyakov \cite{Polyakov:1981rd,Polyakov:1987zb}. If we suppress the contribution to the cosmological constant it is given by,
\beq \label{spol}
S^{pol}_{m} =  -\frac{d}{48\pi} 
\int d^2z d^2z' \sqrt{g}\sqrt{g}'
\left(
 \frac{1}{2}
R(z)\frac{1}{\Delta} R(z')
\right).
\eeq
%
\section{Fixing the gauge}
In the conformal gauge $\tilde{g}_{\alpha\beta}\rightarrow\delta_{\alpha\beta} $, our covariant effective action \eqref{result} reduces to a Liouville action, which is of course not surprising since its decomposition in conformal variables \eqref{ansatz}, \eqref{sexpl} is also of Liouville type, 
\beq \label{smcfg}
S^{cfg}_{m} =-\frac{d}{48 \pi} 
 \int d^2 z  \ 
 \left(
\frac{1}{2} 
\delta^{\alpha \beta} \partial_\alpha \phi \partial_\beta \phi + \mu e^\phi 
\right).
\eeq
From Polyakov's computation of the gravitational part of the path integral measure \cite{Polyakov:1981rd}, which corresponds to the determinant of the $P$ operator \eqref{pop},  we know that the effective action gets an additional contribution such that,
\beq \label{stotcfg}
S^{cfg}_{tot} =
\frac{26 - d}{48 \pi} 
 \int d^2 z  \ 
 \left(
\frac{1}{2} 
\delta^{\alpha \beta} \partial_\alpha \phi \partial_\beta \phi + \mu_{tot} e^\phi 
\right).
\eeq
where we introduced the renormalized cosmological constant $\mu_{tot}$. Since \eqref{smcfg} is just the conformally gauge fixed version of \eqref{result} we propose that the total covariant, non-gauge fixed action corresponding to \eqref{stotcfg} is,
\beq \label{stotf}
S_{tot}=\frac{26 - d}{48 \pi} S_F,
\eeq
where $S_F$ is given by \eqref{result}. Although the covariant form \eqref{stotf} is likely to be true, a gauge independent computation of the determinant in the ghost sector will be performed in a forthcoming publication.

In the synchronous or proper-time gauge, given by 
\beq
g_{\alpha\beta} \rightarrow 
\left(\begin{array}{cc}1 & 0 \\0 & l^2(t,x) \end{array}\right),
\eeq 
the action \eqref{result} reduces to an effective action known from causal dynamical triangulation (CDT) \cite{Ambjorn:1998xu, Ambjorn:2006hu} models,
\beq \label{stotpg}
S^{pg}_{tot} = \frac{26-d}{48\pi}
 \int d^2 z  \ 
 \left(
\frac{2\dot{l}^2}{l}+\mu_{tot}  l
\right).
\eeq
See also \cite{Nakayama:1993we,Fukuma:1993tp} for a discussion on relations between quantum gravity in the proper-time gauge and Euclidean quantum gravity, i.e. DT.

The CDT methods have so far not been detailed enough to determine the pre-factor in the effective action, the reason being an absence of analytical results regarding matter coupled causal dynamical triangulations. This paper represents to the best of our knowledge the first analytical derivation of matter coupled to two dimensional quantum gravity that is compatible with the results of CDT.  
The action \eqref{stotpg} still has an infinite dimensional symmetry left,
\beq
l(x,t) = 
\left(\frac{d x}{d x'}\right)^{-1} l(x',t).
\eeq
We can therefore fix the gauge such that the length variable becomes $x$ independent $l(x,t)\rightarrow L(t)$, where we have absorbed the spatial volume.\footnote{For a more detailed discussion of two dimensional induced gravity in the proper-time gauge we refer to \cite{Banks:1983cu} and \cite{Ambjorn:2006hu}.} This fits with the fact that approximating Liouville field theory by Liouville mechanics, i.e. where the Liouville mode does not depend on $x$, yields exact results.

Using the $x$ independence of the length variable in \eqref{stotpg} we anticipate that the total path integral over $d$ scalar fields and two dimensional metrics modulo diffeomorphisms \eqref{formalPI} reduces to the following quantum mechanical path integral,\footnote{There will however be subtleties in the measure, in fact in standard quantum Liouville theory the central charge barrier arises from subtleties in the measure of the Liouville field \cite{David:1988hj,Distler:1988jt}. The explicit computation of the full path integral and a careful discussion of the measure will be presented in a forthcoming paper.}
\beq
Z = \int \mathcal{D} L
\exp
\left[
-\frac{26-d}{48\pi}
 \int dt \ 
 \left( 
\frac{2\dot{L}^2}{L}+\mu_{tot}  L
\right)
\right],
\eeq
which is compatible with causal dynamical triangulations (CDT).
The quantum mechanical nature is especially clear if we do a variable transformation  $x^2(t)=\frac{26-d}{3\pi}L(t) $ and $\omega^2 = \frac{1}{16}\mu_{tot} $,
\beq
Z = \int \mathcal{D}x 
\exp
\left[-
 \int dt \ 
 \left(
 \frac{1}{2} \dot{x}^2+\omega^2 x^2 
\right)
\right].
\eeq
which is nothing but the quantum mechanical path integral for the harmonic oscillator. 

While the proper-time gauge fixing is in perfect agreement with CDT it seems at odds with quantum Liouville theory and standard matrix model computations for Euclidean dynamical triangulations (DT). This difference has not been completely understood but much progress has been made in the string field theory and matrix model approaches to CDT \cite{Ambjorn:2008gk, Ambjorn:2009rv}. We hope that the current paper will help to clarify the relation even further. So far it has been found that the addition of a coupling constant to suppress so-called baby universes allows one to interpolate between the causal and Euclidean dynamical triangulation phases \cite{Ambjorn:2008gk, Ambjorn:2009rv}. 

A similar transition has been observed in the context of four dimensional computer simulations  \cite{Laiho:2011ya}. The authors incorporate a type of curvature weights to the Euclidean dynamical triangulation model and see both DT and CDT like phases.

There is moreover some reason to believe that the quantum mechanical results of the proper-time gauge calculations might actually be compatible with the ``wrong branch'' of quantum Liouville theory \cite{Ambjorn:2009rv}.
\section{Discussion and outlook}
In this paper we have presented a simple Yang-Mills like action for the integrated Weyl anomaly \eqref{result}. Unlike Polyakov's induced action \eqref{spol} it is local and therefore convenient to work with, especially when comparing different gauges. Although local, our proposed action is not invariant under local Lorentz transformations. We have thus traded non locality for non invariance under local rotations of the gravitational frame vectors (the vielbeins). This implies that the conformal anomaly excites the local rotational degrees of freedom of the gravitational potentials $e^a_\alpha$.

With hindsight such a breaking is not too surprising, since it also occurs in Ho\v{r}ava Lifshitz gravity, and at least in four dimensions intimate connections between CDT and Ho\v{r}ava gravity are known \cite{Ambjorn:2010hu}. For a discussion on two dimensional Ho\v{r}ava-Lifshitz gravity, Einstein-aether theory and their equivalence see \cite{Sotiriou:2011dr}. Our action \eqref{result} is quite similar to the one in \cite{Sotiriou:2011dr}, especially in the Einstein-aether form. The difference is however that our action does not single out a preferred direction in the sense of a distinguished vector field, the two covectors contained in the frame field $e^0_\alpha, e^1_\alpha$ are treated democratically.

An obvious question that arises from our work is whether there exists a local induced action in four dimensions too. We are currently studying this and the possible connection with the computer simulations of four dimensional  CDT \cite{Ambjorn:2010hu}. 

Additionally we wish to develop our ideas into a full fledged covariant formulation of non critical string theory. This is especially interesting since results from CDT indicate that there might be a phase of two dimensional quantum gravity which does not see the central charge barrier of standard quantum Liouville theory, see for instance \cite{Ambjorn:1999yv}.

\ack
The authors would like to thank J. Ambj\o rn for inspirational discussions at an early stage of this work. W.W. also thanks J. Koksma for his support and the Rudolf Peierls Center at Oxford University for providing hospitality. S.Z. would like to thank the STFC under grant ST/G000492/1 for financial support.

\appendix
\section{The Einstein-Hilbert action in two dimensions is a total derivative}
In two dimensions one has the special situation that the group of local rotations is abelian. This can be seen from the fact that only in two dimensions  the generators of the rotation group satisfy the following identity,
\beq
(M^{ab})_{cd}=(\delta^a_c \delta^b_d - \delta^a_d \delta^b_c) =  \epsilon_{cd}\epsilon^{ab}.
\eeq
So one is left with a single generator $\epsilon^{ab}$ which commutes with itself. This implies that the spin connection can be written as 
\beq
\omega_\beta^{ab} = \frac{1}{2} \omega_\beta \epsilon^{ab}, \qquad \omega_\beta = \omega_\beta^{cd}\epsilon_{cd}, 
\eeq
It furthermore has the consequence that the Einstein-Hilbert Lagrangian density in two dimensions can be written as a total derivative 
\begin{eqnarray}
\sqrt{g}R 
&= \tilde{\varepsilon}^{\alpha\beta}\varepsilon^{\gamma\delta}R_{\alpha\beta\gamma\delta}\nonumber = \tilde{\varepsilon}^{\alpha\beta}\epsilon_{ab}R_{\alpha\beta}^{~~~ab}\nonumber\\
&= 2\partial_{\alpha}(\tilde{\varepsilon}^{\alpha\beta} \omega_\beta).
\end{eqnarray}
where we have used the fact that $\epsilon_{ab}\omega_\alpha^{~ac}\omega_{\beta c}^{~~~b} 
= \epsilon_{ab}\delta^{ab}\omega_\alpha \omega_{\beta}=0$. The explicit form of the spin connection is found by solving the zero torsion constraint,
\beq
\partial_{[\alpha} e^a_{\beta]} + \epsilon^a_{~b} \omega_{[\alpha} e_{\beta]}^b = 0,
\eeq
which gives
\beq
\omega_\alpha = -\frac{1}{2}\varepsilon^{\gamma\delta}e_{\alpha a}\partial_{[\gamma} e^a_{\delta]}. 
\eeq

\newpage
\section*{References}



\begin{thebibliography}{99}


\bibitem{Polyakov:1981rd}
  A.~M.~Polyakov,
  ``Quantum geometry of bosonic Strings,''
  Phys.\ Lett.\  {\bf B103 } (1981)  207-210.

\bibitem{Hatfield:1992rz}
  B.~Hatfield,
  ``Quantum field theory of point particles and strings,''
  Redwood City, USA: Addison-Wesley (1992) 734 p. (Frontiers in physics, 75)

\bibitem{Alvarez:1982zi}
  O.~Alvarez,
  ``Theory of strings with boundaries: fluctuations, topology, and quantum geometry,''
  Nucl.\ Phys.\  {\bf B216 } (1983)  125.

\bibitem{O'Raifeartaigh:1996hf}
  L.~O'Raifeartaigh, I.~Sachs, C.~Wiesendanger,
  ``Weyl gauging and curved space approach to scale and conformal invariance,''
  DIAS-STP-96-06, (1996).

\bibitem{Jackiw:2005su}
  R.~Jackiw,
  ``Weyl symmetry and the Liouville theory,''
  Theor.\ Math.\ Phys.\  {\bf 148 } (2006)  941-947.
  [hep-th/0511065].


 
\bibitem{Arcos:2005ec}
  H.~I.~Arcos, J.~G.~Pereira,
  ``Torsion gravity: a reappraisal,''
  Int.\ J.\ Mod.\ Phys.\  {\bf D13 } (2004)  2193-2240.
  [gr-qc/0501017].


\bibitem{Polyakov:1987zb}
  A.~M.~Polyakov,
  ``Quantum Gravity in Two-Dimensions,''
  Mod.\ Phys.\ Lett.\  {\bf A2 } (1987)  893.


\bibitem{Ambjorn:1998xu}
  J.~Ambj\o rn, R.~Loll,
  ``Nonperturbative Lorentzian quantum gravity, causality and topology change,''
  Nucl.\ Phys.\  {\bf B536 } (1998)  407-434.
  [hep-th/9805108]

\bibitem{Ambjorn:2006hu}
  J.~Ambj\o rn, R.~Janik, W.~Westra, S.~Zohren,
  ``The emergence of background geometry from quantum fluctuations,''
  Phys.\ Lett.\  {\bf B641 } (2006)  94-98.
  [gr-qc/0607013]


\bibitem{Nakayama:1993we}
  R.~Nakayama,
  ``2-D quantum gravity in the proper time gauge,''
  Phys.\ Lett.\  {\bf B325 } (1994)  347-353.
  [hep-th/9312158].
  
\bibitem{Fukuma:1993tp}
  M.~Fukuma, N.~Ishibashi, H.~Kawai, M.~Ninomiya,
  ``Two-dimensional quantum gravity in temporal gauge,''
  Nucl.\ Phys.\  {\bf B427}, 139-157 (1994).
  [hep-th/9312175].
  



\bibitem{Banks:1983cu}
 T.~Banks, L.~Susskind,
  ``Canonical quantization of (1+1)-dimensional gravity,''
  Int.\ J.\ Theor.\ Phys.\  {\bf 23 } (1984)  475.

\bibitem{David:1988hj}
  F.~David,
  ``Conformal Field Theories Coupled to 2d gravity in the conformal gauge,''
  Mod.\ Phys.\ Lett.\  {\bf A3 } (1988)  1651.

\bibitem{Distler:1988jt}
  J.~Distler, H.~Kawai,
  ``Conformal field theory and 2d quantum gravity or who's afraid of Joseph Liouville?,''
  Nucl.\ Phys.\  {\bf B321 } (1989)  509.

\bibitem{Ambjorn:2008gk}
  J.~Ambj\o rn, R.~Loll, Y.~Watabiki, W.~Westra, S.~Zohren,
  ``A new continuum limit of matrix models,''
  Phys.\ Lett.\  {\bf B670}, 224-230 (2008).
  [arXiv:0810.2408 [hep-th]].

\bibitem{Ambjorn:2009rv}
  J.~Ambj\o rn, R.~Loll, Y.~Watabiki, W.~Westra, S.~Zohren,
  ``New aspects of two-dimensional quantum gravity,''
  Acta Phys.\ Polon.\  {\bf B40 } (2009)  3479-3507.
  [arXiv:0911.4208 [hep-th]].
  
\bibitem{Laiho:2011ya}
  J.~Laiho, D.~Coumbe,
  ``Evidence for asymptotic safety from lattice quantum gravity,''
  [arXiv:1104.5505 [hep-lat]].

\bibitem{Ambjorn:2010hu}
  J.~Ambj\o rn, A.~G\"{o}rlich, S.~Jordan, J.~Jurkiewicz and R.~Loll,
  ``CDT meets Ho\v{r}ava-Lifshitz gravity,''
  Phys.\ Lett.\  B {\bf 690} (2010) 413
  [arXiv:1002.3298 [hep-th]].
  
\bibitem{Sotiriou:2011dr}
  T.~P.~Sotiriou, M.~Visser, S.~Weinfurtner,
  ``Lower-dimensional Ho\v{r}ava-Lifshitz gravity,''
  [arXiv:1103.3013 [hep-th]].
  
\bibitem{Ambjorn:1999yv}
  J.~Ambj\o rn, K.~N.~Anagnostopoulos and R.~Loll,
  ``Crossing the c = 1 barrier in 2d Lorentzian quantum gravity,''
  Phys.\ Rev.\  D {\bf 61} (2000) 044010
  [arXiv:hep-lat/9909129].
  






  

  
  



  


  











  





\end{thebibliography}
\end{document}